\documentclass[12pt]{article}
\def\dspace{\baselineskip = .30in}

\begin{document}

\title{ Raman spectra of MgB$_{2}$ at high pressure and topological electronic transition }
\author{ K.P. Meletov$^{a,b}$, J. Arvanitidis$^a$, M. P. Kulakovb$^b$,\\
N. N. Kolesnikov$^b$ and  G.A.
Kourouklis$^a$\\ $^a$Physics Division, School of Technology\\
Aristotle University of Thessaloniki \\
Thessaloniki, GR-54006 Greece\\
$^b$ Institute of Solid State Physics,Russian Academy of Sciences,\\
142432  Chernogolovka, Moscow region, Russia}
\date{ }
\maketitle

\dspace \centerline{\bf Abstract} \vspace{.2in} Raman spectra of
the MgB$_{2}$ ceramic samples were measured as a function of
pressure up to 32 GPa at room temperature. The spectrum at normal
conditions contains a very broad peak at $\sim $590 $cm^{-1}$
related to the E$_{2g}$ phonon mode. The frequency of this mode
exhibits a strong linear dependence in the pressure region from 5
to 18 GPa, whereas beyond this region the slope of the
pressure-induced frequency shift is reduced by about a factor of
two. The pressure dependence of the phonon mode up to $\sim$ 5GPa
exhibits a change in the slope as well
as a "hysteresis" effect in the frequency vs. pressure behavior. These singularities in the E$_{2g}$
mode behavior under pressure support the suggestion that MgB$_{2}$ may undergo a pressure-induced topological electronic transition.\\
Pacs numbers 74.25.kc, 62.50.+P, 74.70.Ad
\newpage
\section*{Introduction}
\par The discovery of superconductivity recently in MgB$_{2}$  [1] has initiated a number of studies related to the pressure behavior of the crystalline structure, phonon spectrum and superconductivity transition temperature of this material [2-8]. The high pressure experiments, which traditionally are used to test the structural stability of materials, can play also an important role in the understanding of the superconductivity mechanism. The experimentally observed pressure-induced linear decrease of  $T_c$ [4,6,7] is in general agreement with theoretical estimations based on the BCS theory. Theoretical calculations show that MgB$_{2}$  can be treated as a phonon mediated superconductor with very strong electron-phonon coupling of the in-plane optical E$_{2g}$ phonon mode to the partially occupied planar boron s bands near the Fermi surface [9,10]. The strong coupling contributes considerably to the anharmonicity of the Raman active E$_{2g}$ mode manifested by its very broad lineshape, ranging from 460 $cm^{-1}$ to 660 $cm^{-1}$ according to various calculations [9, 11-13]. The other three phonon modes of MgB$_{2}$  with symmetries B$_{1g}$, A$_{2u}$ and E$_{1u}$ are harmonic and show an insignificant electron-phonon coupling [9].
\par    The first report on Raman scattering in MgB$_{2}$  revealed a broad asymmetric peak at $\sim $ 580 $cm^{-1}$ [14], while subsequent investigations showed the E$_{2g}$ mode frequency near 615-620 $cm^{-1}$ [5,15]. An initial high pressure Raman experiment up to 15 GPa has shown a large linear pressure shift of the E$_{2g}$    phonon frequency [5]. Further extension of the pressure range up to 44 GPa revealed the change in the slope of the linear pressure dependence at $\sim$ 23 GPa for the isotopically pure Mg$^{10}$B$^2$ samples [4]. Similar singularities are observed in the dependence of $T_c$ vs. relative variation of volume, $V/V_0$, [4], which exhibits change in the slope of the linear dependence near the values of $V/V_0$ corresponding to pressures of $\sim$ 20 GPa and $\sim$ 15 GPa for Mg$^{10}$B$^2$ and Mg$^{11}$B$^2$ isotopically pure samples, respectively. This behavior in the pressure dependence of $T_c$ was also observed at  $\sim$9  GPa, for MgB$_{2}$  samples prepared from natural boron isotope mixture [7,8]. Taking into account that the pressure dependence of the lattice parameters of MgB$_{2}$  is smooth and does not show any structural phase transitions in the pressure range up to 40 GPa [2,3], the observed singularities in the pressure dependence of $T_c $ and E$_{2g}$  phonon frequency were supposed to be related to a Lifshitz isostructural topological electronic transition [16].
\par    We have measured the Raman spectra of  MgB$_{2}$ as a function of pressure, up to 32 GPa, at room temperature. The main goal in our experiments was to study carefully the pressure dependence of E$_{2g}$ phonon mode and to examine possible phase transitions in the MgB$_{2}$ system. Despite the fact that there have been published excellent investigations on high pressure Raman scattering by Struzhkin et al. [4] and Goncharov et al. [5], we believe that the results obtained in the present study show new and interesting aspects and complete somehow the study of the pressure behavior of the E$_{2g}$ phonon mode.

\section*{Experimental Details}
\par    Ceramic samples of  MgB$_{2}$ were prepared by direct synthesis from the constituent elements. The initial materials were amorphous boron powder (natural mixture of isotopes, atomic mass 10.811) and pieces of metallic magnesium, both better than 99.9 \% purity.
The stoichiometric weights of materials were placed in molybdenum
crucible and heated up to 1400$^o$ C into the medium-pressure
furnace under Ar-gas pressure of $\sim$ 12 bar followed by
annealing there for an hour. During the heating, the synthesis
of  is believed to occur at $\sim$ 900$^o$ C. The resulting
product was a bronze-color compact material with density $\sim$
2.23 g/cm$^3$ and grain size from 6 to 30 microns. X-ray powder
diffraction pattern of synthesized samples showed the hexagonal
(${\bf a}$=3.086 \AA and${\bf b}$ =3.52 \AA ) to be the main
constituent, with small quantities of MgO and metallic Mg.
\par    Raman spectra were recorded using a triple monochromator (DILOR XY-500) equipped with a CCD liquid-nitrogen cooled detector system. The spectra were taken in the back-scattering geometry by the use of the micro-Raman system comprising by an OLYMPUS microscope equipped with objectives of 100x and 20x magnification and a spatial resolution of $\sim$1.7
$\mu m$ and  $\sim$ 8 $\mu m$, respectively. Small good facetted
bronze-colored grains of MgB$_{2}$  with typical size of $\sim 20
\mu m$ were selected for Raman measurements. The spectral width
of the system was $\sim $ 8 $cm^{-1}$ and the 514.5 nm line of an
$Ar^+$ laser with beam power below 10 mW, measured before the
cell, was used for excitation. Measurements of the Raman spectra
at high pressure were carried out in two independent pressure
cycles using a diamond anvil cell (DAC) of Mao-Bell type [17].
The small chips of the MgB$_{2}$  ceramics selected for their
prominent  E$_{2g}$   -mode Raman signal were loaded into DAC.
The 4:1 methanol-ethanol mixture was used as pressure
transmitting medium and the ruby fluorescence technique was used
for pressure calibration [18]. The E$_{2g}$ phonon frequency was
obtained by fitting a Gaussian function to the experimental peak
after background subtraction.
\section*{Results and Discussion}
\par    The Raman spectra of ceramic samples of MgB$_{2}$ , taken at normal conditions, consist of a broad peak centered near $\sim$ 590 $cm^{-1}$. This frequency value is lower than the earlier reported frequency value of the E$_{2g}$ mode [4, 14, 15]. Probing the ceramic MgB$_{2}$  samples, using high spatial resolution of the micro-Raman system, gave us the possibility to identify small crystalline grains of MgB$_{2}$  whose Raman spectra differ drastically from that of inclusions of MgO or metallic Mg.
\par The Raman spectra of the MgB$_{2}$  for various pressures up to $\sim$29 GPa and at room temperature are shown in Figure 1. The initial spectrum at 1.1 GPa (Fig. 1a) contains the broad $(FWHM\approx250  cm^{-1})$ peak near $\sim600 cm^{-1}$ related to the Raman active E$_{2g}$ mode. The relatively sharp peak near $\sim880 cm^{-1}$ is associated with a methanol-ethanol mixture peak. The intensity of this peak gradually drops with the increase of pressure and vanishes at $\sim$12 GPa upon mixture solidification. When pressure increases the E$_{2g}$    peak shifts to higher energy (Fig. 1b-1f) and somehow broadens, while its Raman intensity does not change noticeably. The release of pressure, down to 1.2 GPa  (Fig. 1g), restores the main features of the initial Raman spectrum.
\par    The pressure dependence of the E$_{2g}$    mode frequency, worked out for various pressure runs, is shown in Figure 2. The open circles show the data for increasing pressure up to $\sim$20 GPa, while the closed circles are related to the decrease of pressure down to $\sim$1.2 GPa. The data marked by open squares are recorded at the subsequent upstroke pressure cycle from $\sim$1.2 GPa to $\sim$32 GPa performed immediately after the release of pressure without disassembling of the DAC. The shaded areas near $\sim$5 GPa and $\sim$18 GPa separate the regions where the pressure behavior of E$_{2g}$   phonon frequency can be fitted to a linear dependence with different slopes $\partial \omega /\partial P$. The largest slope $\partial \omega /\partial  P=18 cm^{-1}$/GPa is found for the region $5 \leq P \leq$ 18 GPa, while for P$>$18 GPa the slope $\partial \omega /\partial  P$ is 6 $cm^{-1}$/ GPa. The most intrigue behavior is observed in the pressure region 1 bar $\sim$ 5 GPa, where the route (open cycles in Fig. 2) of the two  upstroke pressure cycles (new cell loading) differs from the route of the downstroke (solid cycles in Fig. 2) and upstroke (open squares) cycles without the total release of pressure in the DAC. The slopes $\partial \omega /\partial P $ of both routes are slightly different, $\sim$ 7 $cm^{-1}$/GPa for the new loading and $\sim$ 9 $cm^{-1}$/GPa for the recycling routes. Note that the spread out of experimental data on the E$_{2g}$   mode frequency is consistent with the accuracy in the peak position determination, which was found to be close to $\pm $10 $cm^{-1}$.
\par    The pressure dependence of the mE$_{2g}$   ode frequency demonstrates two singularities near $\sim$5 GPa and $\sim$18 GPa. These results are partly correlated with the Raman data obtained by Struzhkin et al., which have been reported a singularity in the slope of the phonon pressure dependence near $\sim$23 GPa for the isotopic pure Mg$^{10}$B$^2$ sample and near $\sim$15 GPa in the pressure dependence of $T_c$ for the isotopic pure Mg$^{11}$B$^2$ sample [4]. Taking into account that the samples in the present investigation were prepared from natural mixture of boron isotopes we think that the singularity near $\sim$18 GPa has the same origin as those observed in [4] for isotopic pure samples. As for the singularity at $\sim$ 5 GPa, it seems to be a new result revealed by recording the spectra for small steps of pressure increase in this interval, a practice which is not apparent in the other studies [4,5,19].
\par    The experimental data for the pressure dependence of the E$_{2g}$   phonon mode are seemingly in contradiction with the X-ray data on MgB$_{2}$ . Although the Raman data show distinct singularities in their pressure dependence, the pressure dependencies of the ${\bf a}$ and ${\bf c}$ parameters of the hexagonal lattice are smooth and do not show any structural phase transition in the pressure region up to 40 GPa [2,3,5]. The possible explanation for this contradiction may be related to the Lifshitz topological electronic transition [16] related to the pressure-induced changes of the topology of the Fermi surface. In such a transition the electron density of states on the Fermi level as well as the electron dynamics possess some peculiar features, which lead to anomalies of the electron thermodynamic and kinetic characteristics. The band structure calculations for the MgB$_{2}$  [9,12] show the splitting of the planar boron $\sigma $ bands along the $\Gamma - A $ line near the Fermi surface, which creates the conditions for a Lifshitz-type transition under the application of high pressure. Tissen et al. [8] have suggested that MgB$_{2}$  undergoes the Lifshitz topological electronic transition to explain the cusp in the pressure dependence of the $T_c$ near 9 GPa. Later the same suggestion has been used to explain the changes in the slopes of the linear pressure dependencies of the E$_{2g}$   phonon frequency and superconducting transition temperature $T_c$ for isotopic pure Mg$^{11}$B$^2$ and Mg$^{10}$B$^2$ samples [4]. We believe that the manifestation of the electronic topological transition in the pressure dependence of the E$_{2g}$   phonon mode is related to the strong electron-phonon coupling of this mode to the planar boron $\sigma $ bands.
\par    Concerning the singularity in the E$_{2g}$   phonon pressure dependence near $\sim$5 GPa we believe that this may be related to some transformation of the initial ceramic material associated with a trend to phase homogenization under high pressure. It seems that the recovered material is more homogeneous as far as its pressure response and the E$_{2g}$   phonon frequency is lower than that of the starting material, therefore an investigation of the $T_c$ for a high pressure treated ceramic sample might be interesting. In addition, the most detailed X-ray data of Prassides et al. [2], related to the pressure dependence of the lattice volume, may include an indication for some singularity near $\sim$ 4.5 GPa. In any case we think that, to clarify this suggestion, further experiments with high quality crystalline samples are necessary.
\par    Finally, we would like to address the difference in the E$_{2g}$   phonon frequency reported in the various Raman studies at normal conditions [4,5,15]. We think that its origin may be related to the difference in the stoichiometry of ceramic samples. For example, a recent publication [20] indicates that the ceramic samples in fact have various stoichiometries, Mg$_{1-x}$B$_2$ with $0 \leq x \leq 0.2$, and superconducting transition temperature $T_c$ varies accordingly from 37 K to 39 K.
\par In conclusion, the pressure dependence of the E$_{2g}$  phonon mode frequency measured as a function of pressure, up to 32 GPa, shows two singularities near $\sim$5 GPa and $\sim$18 GPa. The singularity at $\sim$5 GPa may be related to the homogenization of ceramic samples induced by pressure, while the singularity at $\sim$18 GPa may be related to a Lifshitz topological electronic transition [16].
\newpage
\section*{Acknowledgments}
\par
 The support by the General Secretariat for Research and Technology (GSRT),
  Greece, grant $\Pi E N E \Delta $, 99$E \Delta $/62 is
  acknowledged.
 K. P. M. acknowledges the support by the GSRT, Greece, and the Russian Foundation for Fundamental Research, Russia,
 grant $\# $ 99-02-17555.

\newpage
\section*{References}

\begin{enumerate}
\item J. Nagamatsu, N. Nakagava, T. Muranaka, Y. Zenitani, and J. Akimitsu, Nature {\bf 410}, 63 (2001).
\item K. Prassides, Y. Iwasa, T. Ito, Dam H. Chi, K. Uehara, E. Nishibori, M. Takata, M. Sakata, Y. Ohishi, O. Shinomura, T. Muranaka, and J. Akimitsu, Phys. Rev. {\bf B 64}, 012509 (2001).
\item P. Bordet, M. Mezour, M. Nunez-Regueiro, M. Monteverde, M. D. Nunez-Regueiro, N. Rogado, K. A. Regan, M. A. Hayward, T. He, S. M. Loureiro, and R. J. Cava, Phys. Rev. {\bf B 64}, 172502 (2001).
\item V. V. Struzhkin, A. F. Goncharov, R. J. Hemley, Ho-kwang Mao, G. Lapertot, S. L. Bud'ko, and P. C. Canfield, cond-mat/0106576.
\item A. F. Goncharov, V. V. Struzhkin, E. Gregoryanz, Jingzhu Hu, R. J. Hemley, Ho-kwang Mao, G. Lapertot, S. L. Bud'ko, and P. C. Canfield, Phys. Rev. {\bf B 64}, 100509 (2001).
\item B. Lorenz, R. L. Meng, and G. W. Chu, Phys. Rev. {\bf B 64}, 012507 (2001).
\item S. Deemyad, J. S. Shcilling, J. D. Jorgensen, and D. G. Hinks, cond-mat/0106057.
\item V. G. Tissen, M. V. Nefedova, N. N. Kolesnikov, and M. P. Kulakov, cond-mat/0105475.
\item T. Yildirim, O. Gülseren, J. W. Lynn, C. M. Brown, T. J. Udovic, Q. Huang, N. Rogado, K. A. Regan, M. A. Hayward, J. S. Slusky, T. He, M. K. Haas, P. Khalifah, K. Inumaru, and R. J. Cava, Phys. Rev. Lett. {\bf 87}, 037001 (2001).
\item T. Loa and K. Syassen, cond-mat/0102462.
\item J. Kortus, I. I. Mazin, K. D. Belashchenko, V. P. Antropov, and L. L. Boyer, Phys. Rev. Lett. {\bf 86}, 4656 (2001).
\item J. M. An and W. E. Pickett, Phys. Rev. Lett. {\bf 86}, 4366 (2001).
\item X. Kong, O. V. Dolgov, O. Jepsen, and O. K. Andersen, Phys. Rev. {\bf B 64}, 020501 (2001).
\item K.-P. Bohen, R. Heid, and B. Renker, Phys. Rev. Lett. {\bf 86}, 5771 (2001).
\item J. Hlinka, I. Gregora, J. Pokorny, Plecenik, P. Kus, L. Satrapinsky, and S.Benacka, Phys. Rev. {\bf B 64}, 140503 (2001).
\item I. M. Lifshitz, JETP {\bf 11}, 1130 (1960).
\item A. Jayaraman, Rev. Sci. Instrum. {\bf 57}, 1013 (1986).
\item D. Barnett, S. Block, and G. J. Piermarini, Rev. Sci. Instrum. {\bf 44}, 1 (1973).
\item A. F. Goncharov, V. V. Struzhkin, E. Gregoryanz, H. K. Mao, R. J. Hemley, G. Lapertot, S. L. Bud'ko, P. C. Canfield, and I. I. Mazin, cond-mat/0106258.
\item M. V. Indenbom, L. S. Uspenskaya, M. P. Kulakov, I. K. Bdikin, S. A. Zver'kov, JETP Lett. {\bf 74}, 304 (2001).

\end{enumerate}
\newpage
\section*{Figure Captions}
Fig. 1. Raman spectra of the MgB$_{2}$  for various pressures up to $\sim$29 GPa
and room temperature. Asterisk indicates a methanol-ethanol mixture peak. \\
Fig. 2. Pressure dependence of the frequency of E$_{2g}$   phonon in MgB$_{2}$ . The open (closed) symbols are related to the increase (decrease) of pressure. The shaded areas show the pressure regions where the changes in the slopes of linear pressure shift were observed.  \\
\end{document}